# Better Than Earth

Planets quite different from our own may be the best homes for life in the Universe.

By René Heller

*Origins Institute, McMaster University, Department of Physics and Astronomy, Hamilton (ON) L8S 4M1, Canada*

rheller@physics.mcmaster.ca
www.physics.mcmaster.ca/~rheller

DO WE INHABIT THE BEST O ALL POSSIBLE WORLDS? German mathematician Gottfried Leibniz thought so, writing in 1710 that our planet, warts and all, must be the most optimal one imaginable. Leibniz's idea was roundly scorned as unscientific wishful thinking, most notably by French author Voltaire in his magnum opus, *Candide.* Yet Leibniz might find sympathy from at least one group of scientists – the astronomers who have for decades treated Earth as a golden standard as they search for worlds beyond our own solar system.

Because earthlings still know of just one living world – our own – it makes some sense to use Earth as a template in the search for life elsewhere, such as in the most Earth-like regions of Mars or Jupiter's watery moon Europa. Now, however, discoveries of potentially habitable planets orbiting stars other than our sun – exoplanets, that is – are challenging that geocentric approach.

Over the past two decades astronomers have found more than 1,800 exoplanets, and statistics suggest that our galaxy harbors at least 100 billion more. Of the worlds found to date, few closely resemble Earth. Instead they exhibit a truly enormous diversity, varying immensely in their orbits, sizes and compositions and circling a wide variety of stars, including ones significantly smaller and fainter than our sun. Diverse features of these exoplanets suggest to me and to others that Earth may not be anywhere close to the pinnacle of habitability. In fact, some exoplanets, quite different from our own, could have much higher chances of forming and maintaining stable biospheres. These "superhabitable worlds" may be the optimal targets in the search for extraterrestrial, extrasolar life.

## AN IMPERFECT PLANET

OF COURSE, our planet does possess a number of properties that, at first glance, seem ideal for life. Earth revolves around a sedate, middle-aged star that has shone steadily for billions of years, giving life plenty of time to arise and evolve. It has oceans of life-giving water, largely because it orbits within the sun's "habitable zone," a slender region where our star's light is neither too intense nor too weak. Inward of the zone, a planet's water would boil into steam; outward of the area, it would freeze into ice. Earth also has a life-friendly size: big enough to hold on to a substantial atmosphere with its gravitational field but small enough to ensure gravity does not pull a smothering, opaque shroud of gas over the planet. Earth's size and its rocky composition also give rise to other boosters of habitability, such as climate-regulating plate tectonics and a magnetic field that protects the biosphere from harmful cosmic radiation.

Yet the more closely we scientists study our own planet's habitability, the less ideal our world appears to be. These days habitability varies widely across Earth, so that large portions of its surface are relatively devoid of life – think of arid deserts, the nutrient-poor open ocean and frigid polar regions. Earth's habitability also varies over time. Consider, for instance, that during much of the Carboniferous period, from roughly 350 million to 300 million years ago, the planet's atmosphere was warmer, wetter and far more oxygen-rich than it is now. Crustaceans, fish and reef-building corals flourished in the seas, great forests blanketed the continents, and insects and other terrestrial creatures grew to gigantic sizes. The Carboniferous Earth may have supported significantly more biomass than our present-day planet, meaning that Earth today could be considered less habitable than it was at times in its ancient past.

Further, we know that Earth will become far less life-friendly in the future. About five billion years from now, our sun will have largely exhausted its hydrogen fuel and begun fusing more energetic helium in its core, causing it to swell to become a "red giant" star that will scorch Earth to a cinder. Long before that, however, life on Earth should already have come to an end. As the sun burns through its hydrogen, the temperature at its core will gradually rise, causing our star's total luminosity to slowly increase, brightening by about 10 percent every billion years. Such change means that the sun's habitable zone is not static but dynamic, so that over time, as it sweeps farther out from our brightening star, it will eventually leave Earth behind. To make matters worse, recent calculations suggest that Earth is not in the middle of the habitable

## The Evolution of the Solar Habitable Zone

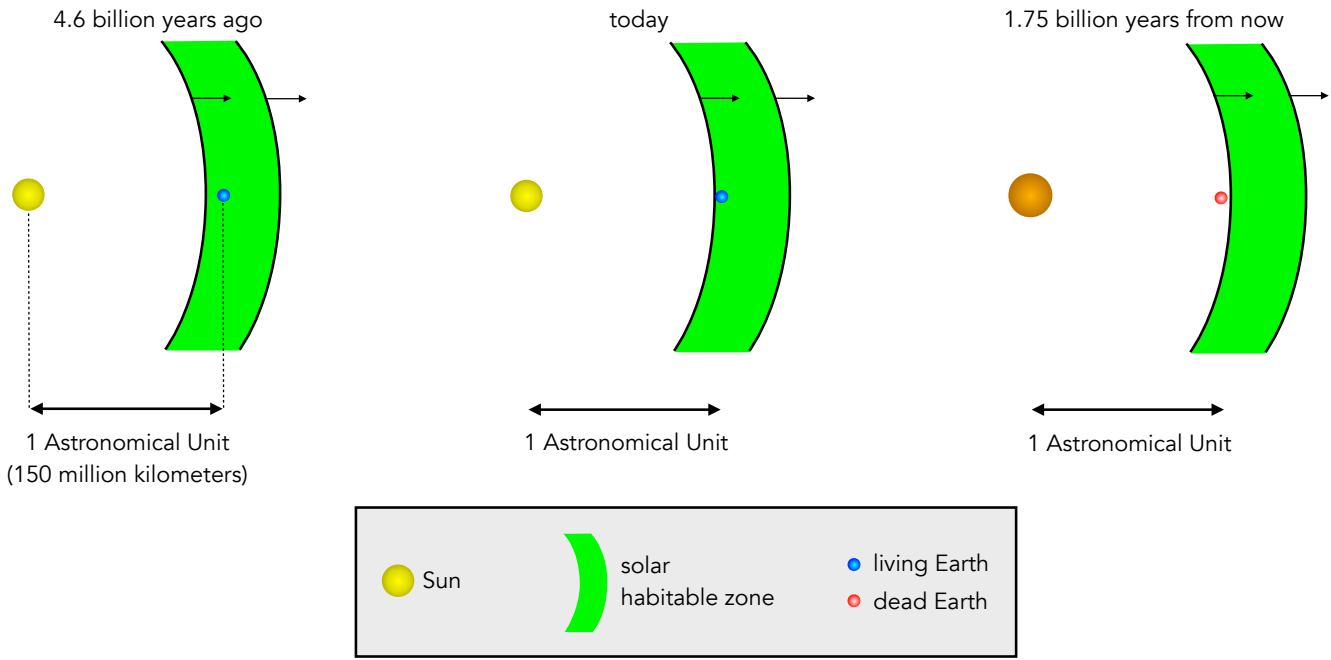

*Figure 1:* This graphic shows the Earth's distance to the Sun and its location in the solar habitable zone (nothing is to scale here) during three different epochs of stellar evolution. In general, the habitable zone of a star is the distance range in which an Earth-like planet would have the ability to sustain liquid surface water – the essential ingredient to make a world habitable. As time goes by (see panel titles), the solar luminosity and radius increase, and so the solar habitable zone moves away from the Sun. In about 1.75 billion years, the Earth will leave the habitable zone and become a desiccated giant rock. While Earth is a nice place to live on today, it can be regarded "marginally habitable" from a cosmological point of view – both in spatial and in temporal dimensions.

zone but rather on the zone's inner cusp, already teetering close to the edge of over-heating (see Figure 1).

Consequently, within about half a billion years our sun will be bright enough to give Earth a feverish climate that will threaten the survival of complex multicellular life. By some 1.75 billion years from now, the steadily brightening star will make our world hot enough for the oceans to evaporate, exterminating any simple life lingering on the surface. In fact, Earth is well past its habitable prime, and the biosphere is fast-approaching its denouement. All things considered, it seems reasonable to say our planet is at present only marginally habitable.

### SEEKING A SUPERHABITABLE WORLD

IN 2012 I FIRST BEGAN THINKING about what worlds more suitable to life might look like while I was researching the possible habitability of massive moons orbiting gas-giant planets. In our solar system, the biggest moon is Jupiter's Ganymede, which has a mass only 2.5 percent that of Earth – too small to easily hang on to an Earth-like atmosphere. But I realized that there are plausible ways for moons approaching the mass of Earth to form in other planetary systems, potentially around giant planets within their stars' habitable zones, where such moons could have atmospheres similar to our own planet.

Such massive "exomoons" could be superhabitable because they offer a rich diversity of energy sources to a potential biosphere. Unlike life on Earth, which is powered primarily by the sun's light, the biosphere of a super-habitable exomoon might also draw energy from the reflected light and emitted heat of its nearby giant planet or even from the giant planet's gravitational field. As a moon orbits around a giant planet, tidal forces can cause its crust to flex back and forth, creating friction that heats

### IN BRIEF

**Astronomers are searching** for twins of Earth orbiting other sunlike stars. **Detecting Earth-like twins** remains at the edge of our technical capabilities. **Larger "super-Earths"** orbiting smaller stars are easier to detect and may be the most common type of planet. **New thinking suggests** that these systems, along with massive moons orbiting gas-giant planets, may also be superhabitable – more conducive to life than our own familiar planet.

the moon from within. This phenomenon of tidal heating is probably what creates the subsurface oceans thought to exist on Jupiter's Europa and Saturn's moon Enceladus. That said, this energetic diversity would be a double-edged sword for a massive exomoon because slight imbalances among the overlapping energy sources could easily tip a world into an uninhabitable state.

No exomoons, habitable or otherwise, have yet been detected with certainty, although some may sooner or later be revealed by archival data from observatories such as NASA's Kepler space telescope. For the time being, the existence and possible habitability of these objects remain quite speculative.

Superhabitable planets, on the other hand, may already exist within our catalogue of confirmed and candidate exoplanets. The first exoplanets found in the mid-1990s were all gas giants similar in mass to Jupiter and orbiting far too close to their stars to harbor any life. Yet as planet-hunting techniques have improved over time, astronomers have begun finding progressively smaller planets in wider, more clement orbits. Most of the planets discovered over the past few years are so-called super-Earths, planets larger than Earth by up to 10 Earth masses, with radii between that of Earth and Neptune. These planets have proved to be extremely common around other stars, yet we have nothing like them orbiting the sun, making our own solar system appear to be a somewhat atypical outlier.

Many of the bigger, more massive super-Earths have radii suggestive of thick, puffy atmospheres, making them more likely to be "mini Neptunes" than super-sized versions of Earth. But some of the smaller ones, worlds perhaps up to twice the size of Earth, probably do have Earth-like compositions of iron and rock and could have abundant liquid water on their surfaces if they orbit within their stars' habitable zones. A number of the potentially rocky super-Earths, we now know, orbit stars called M dwarfs and K dwarfs, which are smaller, dimmer and much longer-lived than our sun. In part because of the extended lives of their diminutive stars, these super-sized Earths are currently the most compelling candidates for super-habitable worlds, as I have shown in recent modeling work with my collaborator John Armstrong, a physicist at Weber State University.

## THE BENEFITS OF LONGEVITY

WE BEGAN OUR WORK with the understanding that a truly longlived host star is the most fundamental ingredient for superhabitability; after all, a planetary biosphere is unlikely to survive its sun's demise. Our sun is 4.6 billion years old, approximately halfway through its estimated 10-billion-year lifetime. If it were slightly smaller, however, it would be a much longerlived K dwarf star. K dwarfs have less total nuclear fuel to burn than more massive stars, but they use their fuel more efficiently, increasing their longevity. The middle-aged K dwarfs we observe today are billions of years older than the sun and will still be shining billions of years after our star has expired. Any potential biospheres on their planets would have much more time in which to evolve and diversify.

A K dwarf's light would appear somewhat ruddier than the sun's, as it would be shifted more toward the infrared, but its spectral range could nonetheless support photosynthesis on a planet's surface. M dwarf stars are smaller and more parsimonious still and can steadily shine for hundreds of billions of years, but they shine so dimly that their habitable zones are very close-in, potentially subjecting planets there to powerful stellar flares and other dangerous effects. Being longer-lived than our sun yet not treacherously dim, K dwarfs appear to reside in the sweet spot of stellar superhabitability.

Today some of these long-living stars may harbor potentially rocky super-Earths that are already several billion years older than our own solar system. Life could have had its genesis in these planetary systems long before our sun was born, flourishing and evolving for billions of years before even the first biomolecule emerged from the primordial soup on the young Earth. I am particularly fascinated by the possibility that a biosphere on these ancient worlds might be able to modify its global environment to further enhance habitability, as life on Earth has done. One prominent example is the Great Oxygenation Event of about 2.4 billion years ago, when substantial amounts of oxygen first began to accumulate in Earth's atmosphere. The oxygen probably came from oceanic algae and eventually led to the evolution of more energy-intensive metabolisms, allowing creatures to have bigger, more durable and active bodies. This advancement was a crucial step toward life's gradual emergence from Earth's oceans to colonize the continents. If alien biospheres exhibit similar trends toward environmental enhancement, we might expect planets around long-lived stars to become somewhat more habitable as they age.

To be superhabitable, exoplanets around small, long-lived stars would need to be more massive than Earth. That extra bulk would forestall two disasters most likely to befall rocky planets as they age. If our own Earth were located in the habitable zone of a small K dwarf, the planet's interior would have grown cold long before the star expired, inhibiting habitability. For example, a planet's internal heat drives volcanic eruptions and plate tectonics, processes that replenish and recycle atmospheric levels of the greenhouse gas carbon dioxide. Without those processes, a planet's atmospheric $CO_2$ would steadily decrease as rainfall washed the gas out of the air and into rocks. Ultimately the $CO_2$-dependent global greenhouse effect would grind to a halt, increasing the likelihood that an Earth-like planet would enter an uninhabitable "snowball" state in which all of its surface water freezes.

Beyond the potential breakdown of a planet-warming greenhouse effect, the cooling interior of an aging rocky world could also cause the collapse of any protective planetary magnetic field. Earth is shielded by a magnetic field generated by a spinning, convecting core of molten iron, which acts like a dynamo. The core remains liquefied because of leftover heat from the planet's formation, as well as from the decay of radioactive isotopes. Once a rocky planet's internal heat reservoir became exhausted, its core would solidify, the dynamo would cease, and the magnetic shield would fall, allowing cosmic radiation and stellar flares to erode the upper atmosphere and impinge on the surface. Consequently, old Earth-like planets would be expected to lose substantial portions of their atmospheres to space, and higher levels of damaging radiation could harm surface life.

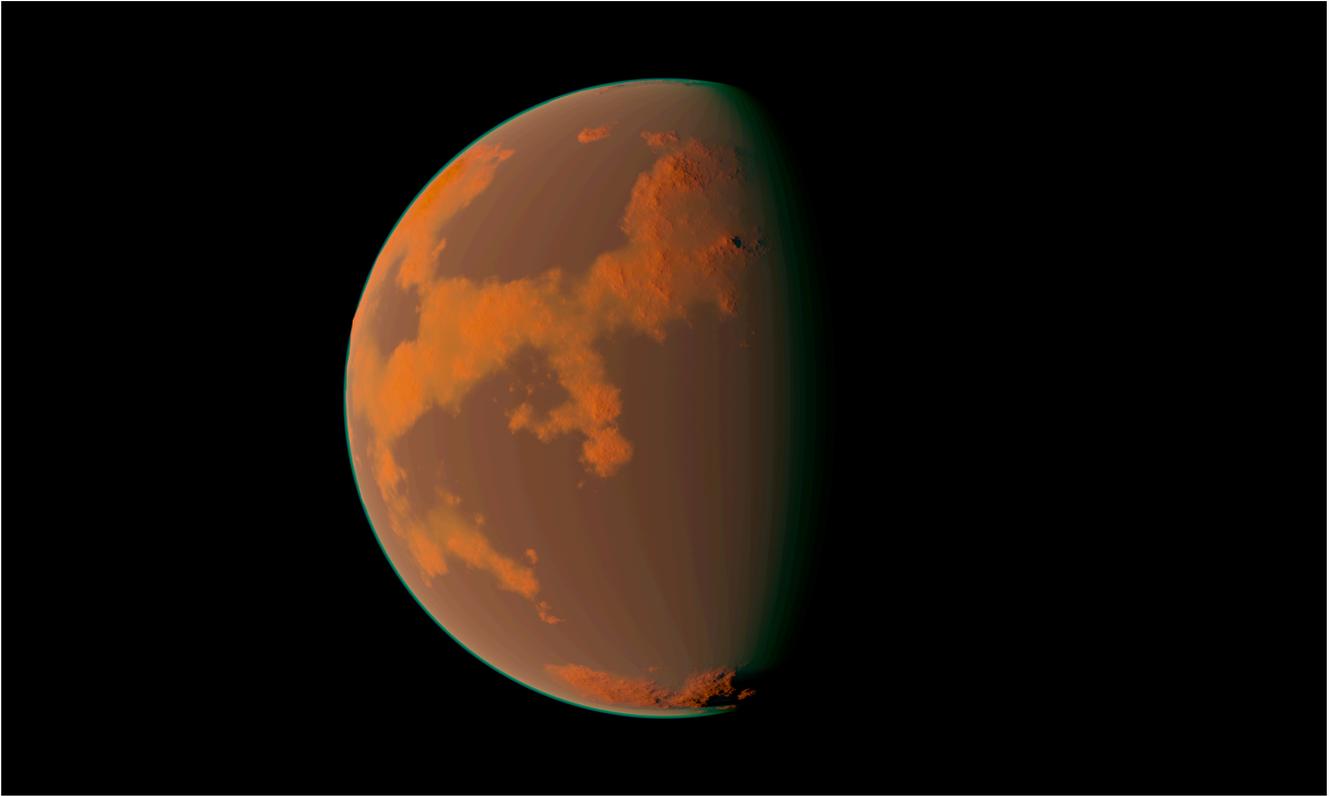

*Figure 2:* *This example of a hypothetical superhabitable planet visualizes a range of effects that we expect these class of planets to show. First, it appears illuminated by an orange light source, which is owed to its host star being a so-called K dwarf star, a star about 80 percent the mass of the Sun. Second, the planet shows a somewhat opaque atmosphere, which is due to its increased surface gravity compared to Earth and its ability to draw down more gas). Third, while this planet has a large amount of water, its land surface is somewhat more dispersed than on Earth. It resembles an archipelagos world rather than a planet dominated by continents, like Earth.*

Rocky super-Earths as much as twice our planet's size should age more gracefully than Earth, retaining their inner heat for much longer because of their significantly greater bulks. But planets larger than about three to five Earth masses may actually be too bulky for plate tectonics because the pressures and viscosities in their mantles become so high that they inhibit the required outward flow of heat. A rocky planet only two times the mass of Earth should still possess plate tectonics and could sustain its geologic cycles and magnetic field for several billion years longer than Earth could. Such a planet would also be about 25 percent larger in diameter than Earth, giving any organisms about 56 percent more surface area than our world on which to live.

### LIFE ON A SUPERHABITABLE SUPER-EARTH

WHAT WOULD A SUPERHABITABLE PLANET look like? Higher surface gravity would tend to give a middling super-Earth planet a slightly more substantial atmosphere than Earth's, and its mountains would erode at a faster rate. In other words, such a planet would have relatively thicker air and a flatter surface. If oceans were present, the flattened planetary landscape could cause the water to pool in large numbers of shallow seas dotted with island chains rather than in great abyssal basins broken up by a few very large continents (see Figure 2). Just as biodiversity in Earth's oceans is richest in shallow waters near coastlines, such an "archipelago world" might be enormously advantageous to life. Evolution might also proceed more quickly in isolated island ecosystems, potentially boosting biodiversity.

Of course, lacking large continents, an archipelago world would potentially offer less total area than a continental world for land-based life, which might reduce overall habitability. But not necessarily, especially given that a continent's central regions could easily become a barren desert as a result of being far from temperate, humid ocean air. Furthermore, a planet's habitable surface area can be dramatically influenced by the orientation of its spin axis with respect to its orbital plane around its star. Earth, as an example, has a spin-orbit axial tilt of about 23.4 degrees, giving rise to the seasons and smoothing out what would otherwise be extreme temperature differences between the warmer equatorial and colder polar regions. Compared with Earth, an archipelago world with a favorable spin-orbit alignment could have a warm equator as well as warm, ice-free poles and, by virtue of its larger size and larger surface area on its globe, would potentially boast even more life-suitable land than if it had large continents.

Taken together, all these thoughts about the features important to habitability suggest that superhabitable worlds are slightly larger than Earth and have host stars somewhat smaller and dimmer than the sun. If correct, this conclusion is tremendously exciting for astronomers because across interstellar distances super-Earths orbiting small stars are much easier to detect and study than twins of our own Earth-sun system. So far statistics from exoplanet surveys suggest that super-Earths around small stars are substantially more abundant throughout our galaxy than Earth-sun analogues. Astronomers seem to have many more tantalizing places to hunt for life than previously believed.

One of Kepler's prize finds, the planet Kepler-186f, comes to mind. Announced in April 2014, this world is 11 percent larger in diameter than Earth and probably rocky, orbiting in the habitable zone of its M dwarf star. It is probably several billion years old, perhaps even older than Earth. It is about 500 lightyears away, placing it beyond the reach of current and near-future observations that could better constrain predictions of its habitability, but for all we know, it could be a superhabitable archipelago world.

Closer superhabitable candidates orbiting nearby small stars could soon be discovered by various projects, most notably the European Space Agency's PLATO mission, slated to launch by 2024. Such nearby systems could become prime targets for the James Webb Space Telescope, an observatory scheduled to launch in 2018, which will seek signs of life within the atmospheres of a small number of potentially superhabitable worlds. With considerable luck, we may all soon be able to point to a place in the sky where a more perfect world exists.

**The author**: René Heller is a postdoctoral fellow at the Origins Institute at McMaster University in Ontario and a member of the Canadian Astrobiology Training Program. His research focuses on the formation, orbital evolution, detection and habitability of extrasolar moons. He is informally known as the best German rice pudding cook in the world.

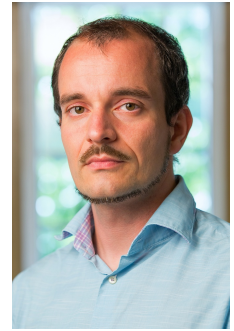